\input harvmac.tex
\noblackbox

\lref\JCardy{J. Cardy, Nucl. Phys. B324 (1989) 581.}
\lref\Burk{T. W. Burkhardt, T. Xue, Nucl. Phys. B354 (1991) 653.}
\lref\Chakrev{A.J. Leggett, S. Chakravarty, A.T. Dorsey,
M.P.A. Fisher, A. Garg, W. Zwerger, Rev. Mod. Phys. 59, (1987) 1.}
\lref\Smir{F. Smirnov, ``Form-factors in completely integrable
models of quantum field theory'',
Worls Scientific (Singapore) 1990.}
\lref\GZ{S. Ghoshal, A.B. Zamolodchikov, Int. J. Mod. Phys.
A9, (1994) 3841.}
\lref\Egger{R. Egger, H. Grabert, Phys. Rev. B55 (1997) R3809.}
\lref \FLS{P. Fendley, F. Lesage, H. Saleur, J. Stat. Phys.
85, (1996) 211.}
\lref\LSS{F. Lesage, H. Saleur, S. Skorik, Nucl. Phys. B474,
(1996) 602.}
\lref\BPZ{A.A. Belavin, A.M. Polyakov, A.B. Zamolodchikov,
Nucl. Phys. {\bf B}241, (1984) 333.}

\Title{USC-98-001}
{\vbox{
\centerline{Boundary conditions changing operators}
 \vskip 4pt
\centerline{in non conformal theories.}}}

\centerline{F. Lesage and H. Saleur\footnote{$^\dagger$}
{Packard Fellow}}
\bigskip\centerline{Department of Physics}
\centerline{University of Southern California}
\centerline{Los Angeles, CA 90089-0484}

\vskip .3in
Boundary conditions changing operators have played an important role
in
conformal field theory. Here, we study their equivalent in
the case where a mass scale is introduced, in an integrable way,
either in the bulk or at the boundary.
More precisely, we propose an axiomatic approach to
determine the general scalar products ${}_b\langle
\theta_1,\ldots,\theta_m||\theta'_1,\ldots,\theta'_{n}\rangle_a$
between asymptotic states in the Hilbert spaces with $a$ and $b$
boundary conditions respectively, and compute these scalar products
explicitely in the case of the Ising and sinh-Gordon models
with a mass and a boundary interaction.
These quantities can be used to study statistical
systems with inhomogeneous boundary conditions, and, more
interestingly maybe, dynamical problems in quantum impurity problems.
As an example, we obtain a series of new exact results
for  the transition probability  in the
double well problem of dissipative quantum mechanics.

\Date{01/98}

\newsec{Introduction}

The concept of boundary conditions changing operators has been
of crucial importance in the analysis of boundary conformal field
theories, as well as in their applications to quantum impurity
problems.  In the last few years, important progress has been made
in extending the solution of conformal field theories to theories
perturbed either by a bulk or by  a boundary operator.
The latter have extremely interesting
applications to the study of flows in quantum impurity problems.

As we will show here, it is possible to introduce boundary
conditions changing operators even in the case where there
is a mass scale, either in the bulk or at the
boundary. It is not completely clear what the formal use of these
objects (eg as illustrated in \JCardy\ for the conformal case)
might be, but they certainly do have applications.
{}From a 2D, statistical mechanics,
point of view, one might wonder what is the effect of having
different parts of the boundary with different boundary
conditions \Burk, describing, for instance
a situation with a (classical)  ``impurity'' on the boundary.
{}From the 1+1 point of view of quantum impurity problems, one might
wish to describe situations where the coupling to the impurity
is changed at some particular time, and one is
interested in the subsequent time evolution of the degrees of
freedom. In fact, one of the key observables in the two state
problem of dissipative quantum
mechanics, the well known quantity  ``$P(t)$'' (see below), is
essentially defined in that fashion \Chakrev.

In the conformal situation, only a discrete set of conformal boundary
conditions are available, and the effect of switching from one
to the other is described by the insertion on the boundary of the
appropriate operator. For instance, in the Ising
model, one goes from free to fixed spins by inserting the
conformal operator $\Phi_{12}$ of weight $h={1\over 16}$,
and from fixed up to fixed down by inserting
$\Phi_{13}$ of weight $h={1\over 2}$ \JCardy (where the $\Phi_{rs}$
are the usual degenerate conformal fields \BPZ).
We are interested in the more general case where conformal
invariance is broken at the boundary, and also
maybe in the bulk. As an example, we can consider a situation
where, in the Ising model, we apply a boundary magnetic field
$h_a$ for $y<0$, $h_b$ for $y>0$ (we use coordinates $x$ and $y$
to describe the two dimensional space and the boundary sits
at $x=0$ in this paper), and in addition, $T\neq T_c$ in
the bulk. In general, one does not expect this situation to be
described by the insertion of a simple operator at $y=0$ anymore.
Nevertheless, the change of boundary conditions
can be fully characterized in the factorized scattering description
(the Ising model with a boundary field being integrable) by scalar
products of asymptotic states in the Hilbert space with $h_a$ and
$h_b$ (see below), and the ``operator'' inserted at $y=0$
can  thus be written, in principle,  in terms of the Faddeev
Zamolodchikov algebra. The problem then reduces to the determination
of scalar products (we also call them transition factors),
and is similar in nature to the problem of determining form-factors
in integrable theories \Smir.

In this paper, we mostly restrict to the Ising case, where the
computations are already moderately complicated, and obtain
the complete solution to the problem with changing boundary
magnetic fields in the massive theory.
In section 2, we discuss the general problem of form-factors
in the crossed channel. We determine the
scalar products of interest for $h_ah_b>0$ in section 3. In
section 4, we discuss the limit where the bulk is massless,
including the conformal invariant case.  In section 5, we discuss
the case $h_ah_b<0$ which require a different treatment.
Section 6 is devoted to extending some of these results to
the sinh-Gordon model, where the non trivial bulk scattering matrix
introduces
further complications.
In section 7,  based on some reasonable conjectures, we
apply  and generalize our results to the determination of
$P(t)$ in dissipative quantum mechanics, for arbitrary
value of the dissipation parameter $g$.

Some of the results presented in this paper  have appeared in a
shorter
version \ref\shorty{F. Lesage, H. Saleur, ``Boundary interactions
changing operators and dynamical correlations in quantum impurity
problems'', cond-mat/9712019.}.

\newsec{Form factors in the crossed channel}

Consider the massive Ising model defined in the half plane
$x\in (-\infty,0]$, $y\in (-\infty,\infty)$, in the presence
of a boundary magnetic field. The action reads
\eqn\act{A=\int_{-\infty}^0 dx\int_{-\infty}^\infty dy \
a_{FF}(x,y)+{1\over 2}
\int_{-\infty}^\infty
dy\left[\left(\psi\bar{\psi}\right)(x=0)+a\dot{a}\right]
+h\int_{-\infty}^\infty dy\sigma_B(y).}
Here $a_{FF}$ is  the usual massive free Majorana fermion action,
$a$ is a boundary fermion
satisfying $a^2=1$, $\sigma_B$ is the boundary spin operator, which
coincides with ${1\over 2}\left(\psi+\bar{\psi}\right)a$.

As discussed in \GZ\ the problem can be studied
from the point of view of the direct channel with
$x$ taken as time and $y$ as space.  In that case the Hilbert space
is the usual one, and the boundary represented by a boundary state.
On the other hand, in the crossed channel, one has a new Hilbert
space for the theory on the half line.
Setting $z=x+iy$,  the one point function of the energy
is easy to compute in the direct channel. Using the boundary state
formula \GZ\
\eqn\bdrsta{|B\rangle=\exp\left[\int_{-\infty}^\infty
{d\theta\over 4\pi}
K(\theta)Z^*(-\theta)Z^*(\theta)\right]|0\rangle,}
where $K$ is related to the reflection matrix
through $K(\theta)=R\left({i\pi\over 2}-\theta\right)$, and the
two particle (bulk) form factor\foot{Here, our normalization for
bulk asymptotic states is
is $\langle \theta_1|\theta_2\rangle=
2\pi\delta(\theta_1-\theta_2)$.}
\eqn\ffbulk{\langle 0|\epsilon|\theta_1,\theta_2\rangle=
i m \sinh\left({\theta_1-\theta_2\over 2}\right),}
one obtains\foot{Rapidity integrals, unless specified,
run from $-\infty$ to $\infty$.}
\eqn\oneptfctmas{\langle \epsilon(z,\bar{z})\rangle
=-i{m\over 2}\int {d\theta\over 2\pi} K(\theta)
\sinh\theta e^{2mx\cosh\theta}.}
An expression in the crossed channel follows by moving the contour
of integration in the variable $\theta$ ($w=iz$)
\eqn\pfou{\langle \epsilon(w,\bar{w})\rangle
={m\over 2}\int {d\theta\over 2\pi}
R(\theta)\cosh\theta e^{-2imx\sinh\theta}.}
There is no simple way  to compute
this expression directly in the crossed channel, except by solving
the problem explicitely, writing mode expansions for the
fermions (see appendix A). In
particular,  \pfou, which can be considered as a zero particle
form-factor, does not, as far as we know,  follow
from any known nice form of the ground state of the theory with
boundary.

Using the mode decomposition of  fermions, it is possible to
construct form factors with
more particles. Because of the boundary, the states at $\theta$ and
$-\theta$ are related: one has
$Z^*(\theta)|0\rangle_a=R(\theta)Z^*(-\theta)|0\rangle_a$ (here
$|0\rangle_a$
denotes the ground state with boundary condition of type $a$ at
$x=0$, normalized such that ${}_a\langle 0|0\rangle_a=1$. We drop the
label $a$ when it is not necessary).
We denote by $||\theta\rangle_a$ the one particle asymptotic
state on the half line $Z^*(\theta)|0\rangle_a$
\foot{The normalization is the same as in the bulk:
$\langle\theta_1||\theta_2\rangle=2\pi\delta(\theta_1-\theta_2)$.}.
Then one has
\eqn\twopffi{\eqalign{
\langle 0|\epsilon(w,\bar{w})||\theta_1\theta_2\rangle
&=im\  e^{-my(\cosh\theta_1+\cosh\theta_2)}
\cr & \times \prod_{k=1,2} \left[
1+R(\theta_k) t_k\right]
\sinh\left({\theta_1-\theta_2\over 2}\right)
e^{im(\sinh\theta_1+\sinh\theta_2)x}
}}
where $t_i$ is an operator acting on functions of many variables
as follows
\eqn\tidef{
t_i g(\cdots \theta_i \cdots)=g(\cdots -\theta_i \cdots) , \ \
t_i^2=1, \ \
t_i t_j=t_j t_i.
}
This expression, call it ${\cal F}_{aa} (\theta_1\theta_2)$,
obeys the relations
\eqn\rel{{\cal F}_{aa}
 (\theta_1,\theta_2)=
R(\theta_1){\cal F}_{aa} (-\theta_1,\theta_2)=R(\theta_2)
{\cal F}_{aa} (\theta_1,
-\theta_2)=-{\cal F}_{aa}(\theta_2,\theta_1).}
This will allow the use of integrals running from $-\infty$
to $\infty$ in the computation of correlators.

Observe that \twopffi\ does not have a pole at
$\theta_2=\theta_1+i\pi$,
as is only natural for the energy operator. Therefore, in the
crossed channel, there is no relation between the  two particle
and the zero particle form-factor of the energy via
kinematic poles.

In order to better understand and generalize these form factors,
let us follow an axiomatic approach to try to compute
them.  Recently, axioms for form factors in the cross-channel
in the presence of a boundary have been written for the
XXZ spin chain \ref\miwa{M. Jimbo, R. Kedem, H. Konno, T. Miwa,
R. Weston, Nucl. Phys. {\bf B}448 (1995) 429; hep-th/9502060.}.
For notations sake, we define
the form factor in the cross-channel with boundary conditions
of type $a$ to be
\eqn\ffdef{
{\cal F}_{aa}(\theta_1,...,\theta_n)
={}_a\langle 0 |{\cal O}Z^*(\theta_1),\cdots ,
Z^*(\theta_n)|0\rangle_a.
}
Following the authors of \miwa\ and taking
the continuum limit of their axioms we find, in our specific
case that
\eqn\ffcross{
{\cal F}_{aa}(\theta_1,\cdots, \theta_i,\theta_{i+1},\cdots
\theta_n)=
S(\theta_i-\theta_{i+1}) {\cal F}_{aa}(\theta_1,\cdots,
\theta_{i+1},\theta_i,\cdots \theta_n)
}
\eqn\ffdpi{\eqalign{
{\cal F}_{aa}(\theta_1,&\cdots, \theta_i+2\pi i,\cdots \theta_n)=
{\cal F}_{aa}(\theta_1,\cdots, \theta_i,\cdots \theta_n)
S(\theta_{i-1}-\theta_i) \cdots S(\theta_1-\theta_i)
\cr & \times
R_a^*(\theta_i+i\pi) S(-\theta_1-\theta_i)\cdots
S(-\theta_{i-1}-\theta_i) S(-\theta_{i+1}-\theta_i)
\cdots S(-\theta_n-\theta_i)  \cr & \times
R^*_a(\theta_i)S(\theta_n-\theta_i) \cdots
S(\theta_{i+1}-\theta_{i})
}}
The kinematic pole equation gives the residue of the form-factor
as $\theta_j=\theta_1+i\pi$, and it is given by
\eqn\kinres{\eqalign{
&{\rm Res}  {\cal F}_{aa}(\theta_1 ,\cdots,\theta_i,\cdots
,\theta_n)=
i {\cal F}_{aa}(\hat{\theta_1}\cdots \hat{\theta_j}\cdots \theta_n)
\cr &
\left[ S(\theta_1-\theta_{j}) \cdots S(\theta_{j-1}-\theta_j)
- S(\theta_j-\theta_{j+1})
\cdots S(\theta_j-\theta_n)R_a(\theta_j)\right.\cr
&\left.\times S(\theta_n+\theta_j)\ldots S(\theta_2+\theta_j)
R_a^*(\theta_j)\right].\cr
}}
Other residues are also generated from the bound states of the
$S$ and $R$ matrices, if any.

We have written the axioms without internal  indices, which are
implicit in the previous equations. In
the case of diagonal scattering, the $R$ matrix actually drops
out from these equations due to $RR^*=1$. The only
dependence appears in the last, reflection equation
\eqn\reflec{\eqalign{
{\cal F}_{aa}(\theta_1,\cdots ,\theta_i,\cdots ,\theta_n)=&
S(\theta_i-\theta_{i+1}) \cdots S(\theta_i-\theta_n) R_a(\theta_i)
S(-\theta_n-\theta_i)
\cr & \times  \cdots S(-\theta_{i+1}-\theta_{i})
{\cal F}_{aa}(\theta_1,\cdots,-\theta_i,\cdots,\theta_n).
}}
Up until now, solutions to these axioms have not been
written and this is the task we plan to undertake here.
In fact, we will first generalise these form factors
slightly, and this will lead to a more general set of
axioms.

\newsec{The case of mixed boundary conditions}

\subsec{The two particle case}

We now consider  the case of inhomogeneous  boundary:
we suppose that  the boundary conditions for $y\in (-\infty,0]$
are described by a boundary magnetic field $h_a$,
and for $y\in [0,\infty)$ by a field $h_b$.
Suppose we wish to compute again  $\langle \epsilon(
w,\bar{w})\rangle$ in the cross-channel (in the
direct channel, no expression for the boundary state
$|B\rangle$ is available). To do so, we have to be careful
that the asymptotic states will depend on the boundary conditions
in a more intricate  way than through the
simple relation $||\theta\rangle_a=R_a(\theta)||-\theta\rangle_a$.
 Indeed, the scalar products
$$
{}_b\langle \theta_n,\ldots,\theta_1
||\theta_{n+1}\ldots\theta_{n+m}\rangle_a
$$
{\bf have} to be non zero in general, even for disjoint sets of
rapidities ($m$ even), since the asymptotic states provide a
complete set of states for any given boundary condition.
As a result, the one-point function
of the energy with mixed boundary conditions will read (for
$y>0$)
\eqn\newoneptf{\eqalign{
\langle\epsilon(w,\bar{w})\rangle_{ba}=&\int_0^\infty
{d\theta_1d\theta_2\over
8\pi^2}\left\{  {}_b\langle 0||\epsilon(w,\bar{w})||
\theta_1\theta_2\rangle_b\
_b\langle\theta_2\theta_1||0\rangle_a\right.\cr
&\left.+\ _b\langle 0||\epsilon(w,\bar{w})||0\rangle_b\
_b\langle 0||0\rangle_a\right\}/_b\langle 0||0\rangle_a.\cr}}
Let us now introduce the quantities
\eqn\defi{{\ _b\langle\theta_2\theta_1||0\rangle_a\over
\ _b\langle 0||0\rangle_a}=G(\theta_1,\theta_2),}
and
\eqn\defi{{\ _b\langle 0||\theta_1\theta_2\rangle_a\over
\ _b\langle 0||0\rangle_a}=F(\theta_1,\theta_2).}
To determine these scalar products we sometimes call transition
factors, we use, as in the
well known case of operators
in the bulk, an axiomatic approach \Smir.  It is  then
sraightforward to write the first set of axioms
\eqn\axiom{G(\theta_1,\theta_2)=-G(\theta_2,\theta_1)=
R^*_b(\theta_1) G(-\theta_1,\theta_2)
=R^*_b(\theta_2)G(\theta_1,-\theta_2).}
To complement these, we need a residue condition. It is easily
obtained as follows. Consider the one point function of the energy.
It does have different expressions for $y>0$, where
$$
\eqalign{\langle \epsilon\rangle_{ba} = {m\over 2}  \int
{d\theta\over 2\pi}
R_b(\theta)\cosh\theta
e^{-2imx\sinh\theta}-im\int
{d\theta_1d\theta_2\over 8\pi^2}
R_b(\theta_1)\sinh{\theta_1+\theta_2\over 2}\cr
G(\theta_1,\theta_2)
\exp[im(-\sinh\theta_1+\sinh\theta_2)x]
 \exp[-m(\cosh\theta_1+\cosh\theta_2)y]\cr}
$$
and for $y<0$, where
$$
\eqalign{\langle \epsilon\rangle_{ba} = {m\over 2}  \int
{d\theta\over 2\pi}
R_a(\theta)\cosh\theta
e^{-2imx\sinh\theta}
+im\int {d\theta_1d\theta_2\over 8\pi^2}
R_a^*(\theta_1)\sinh{\theta_1+\theta_2\over 2}\cr
F(\theta_1,\theta_2)
\exp[im(\sinh\theta_1-\sinh\theta_2)x]
 \exp[m(\cosh\theta_1+\cosh\theta_2)y],\cr}
$$
where we used the symmetry relations \axiom\ to write all integrals
on the interval $(-\infty,\infty)$\foot{When doing so, one has to
use the normalization $\langle\theta_1||\theta_2\rangle=2\pi\left[
\delta(\theta_1-\theta_2)\right.$$+\left.
R^*(\theta_1)\delta(\theta_1+\theta_2)\right]$.
The presence of two delta terms is compensated by factors ${1\over
2}$
one has to introduce to get the integrations running over the
whole real axis.}. By requiring that the
second expression
is the analytic continuation of the first, we find, moving the
contours of integration,
that $G$ must have a simple pole for $\theta_2=\theta_1-i\pi$, with
residue
as $\theta_2\to \theta_1-i\pi$,
\eqn\residucond{\hbox{Res } G(\theta,\theta-i\pi)= -i \left(1
-{R_a(\theta)\over
R_b(\theta)}\right).}
This is equivalent (see  below) to having a simple pole as
$\theta_2\to\theta_1+i\pi$
\eqn\residucondi{\hbox{Res } G(\theta,\theta+i\pi)= -i \left(1
-{R_b(\theta)\over
R_a(\theta)}\right).}
Similarly, $F$ has poles for the same values $\theta_2=\theta_1\pm
i\pi$, with residues switched.

Recall that the $R$ matrix reads \GZ\
\eqn\rmat{R_a=i \tanh \left(i{\pi\over 4}-{\theta\over 2}\right)
{\kappa_a-i\sinh\theta\over \kappa_a+i\sinh\theta},}
with $\kappa_a=1-h_a^2/(4\pi)$.
It obeys the relations
\eqn\identc{R(\theta+i\pi)=-{1\over R(\theta)}=-R^*(\theta).}
In the following, we parametrize $\kappa_a=-\cosh\theta_a$ (and
similarly for $\kappa_b$).
It is useful to observe that
\eqn\obsa{{\kappa_a-i\sinh\theta\over
\kappa_a+i\sinh\theta}=i\tanh\left({\theta+\theta_a\over
2}-i{\pi\over
4}\right)
i\tanh\left({\theta-\theta_a\over 2}-i{\pi\over 4}\right),}
Let us now try to determine the functions $F,G$. To do so,
consider the equation
\eqn\baseq{
\Phi(\theta)\Phi(\theta+i\pi)={1\over
\cosh \left({\theta-\theta_{b}\over 2}-{i\pi\over 4}\right)
\cosh \left({\theta-\theta_a \over 2}+{i\pi\over 4}\right)}.}
Its simplest solution is
\eqn\defiiii{\eqalign{&\Phi(\theta|\theta_{b},\theta_a)
\equiv {1\over \cosh\left(
{\theta-\theta_{b}\over 2}-{i\pi\over 4}\right)}
\prod_{n=0}^\infty
{\Gamma \left({5\over 4}+n-i{\theta-\theta_{b}\over 2\pi}\right)\over
\Gamma \left({3\over 4}+n-i{\theta-\theta_{b}\over 2\pi}\right)}\cr
&{\Gamma \left({5\over 4}+n+i{\theta-\theta_{b}\over
2\pi}\right)\over
\Gamma \left({3\over 4}+n+i{\theta-\theta_{b}\over 2\pi}\right)}
{\Gamma \left({3\over 4}+n-i{\theta-\theta_a\over 2\pi}\right)\over
\Gamma \left({5\over 4}+n-i{\theta-\theta_a\over 2\pi}\right)}
{\Gamma \left({3\over 4}+n+i{\theta-\theta_a\over 2\pi}\right)\over
\Gamma \left({5\over 4}+n+i{\theta-\theta_a\over 2\pi}\right)},\cr}}
or, using an integral representation,
\eqn\intrep{\Phi(\theta|\theta_{b},\theta_a)\equiv
{1\over \cosh\left(
{\theta-\theta_{b}\over 2}-{i\pi\over 4}\right)}
\exp\left[\int_{-\infty}^\infty {dt\over t}
{e^{i{\theta-\theta_a\over 2\pi}t}
-e^{i{\theta-\theta_{b}\over 2\pi}t}
\over 4\cosh {t\over 4}\sinh{t\over 2}}\right].}
The function $\Phi$  has no poles in the physical strip $\hbox{Im
}\theta\in [0,\pi]$, and a single pole
$\theta=\theta_b-{i\pi\over 2}$ in $\hbox{Im }\theta\in [-\pi,0]$.
It satisfies the following  identities
\eqn\identa{\Phi^*(\theta|\theta_{b},\theta_a)=
\Phi(-\theta|-\theta_{b},-\theta_a)=
{1\over  i\tanh\left({\theta-\theta_{b}\over 2}-i{\pi\over 4}\right)}
\Phi(\theta|\theta_{b}\theta_a).}
together with
\eqn\identb{\Phi(\theta+i\pi|\theta_{b},\theta_a)={1\over i \tanh
\left({\theta-\theta_a\over 2}-i{\pi\over 4}\right)}
\Phi(\theta|\theta_a,\theta_{b}).}
The asymptotics of $\Phi$ can be worked out easily from
the integral representation. If $\theta-\theta_{a,b}\to\infty$,
one has $\Phi\approx 2\omega e^{-\theta/2}
e^{(\theta_a+\theta_b)/4}$
($\omega=e^{i\pi/4}$), while in the
opposite case $\theta-\theta_{a,b}\to -\infty$, one has
$\Phi\approx 2\omega^{-1} e^{\theta/2}e^{-(\theta_a+\theta_b)/4}$.

We now introduce the quantity
\eqn\fundsolmas{f(\theta|\theta_{b},\theta_a)
=\sqrt{-i(\kappa_a-\kappa_b})
\Phi(\theta|\theta_{b},\theta_a)
\Phi(\theta|-\theta_{b},-\theta_a).}
It is important to stress that, despite the notation, this function
is
independently an even function of $\theta_a$ and $\theta_b$. By
convention,
we always chose these variables to have  positive real part in what
follows. The function $f$  obeys
\eqn\identd{\eqalign{f(\theta)f(\theta+i\pi)=
&{(\kappa_a-\kappa_b)\over
i\cosh\left( {\theta+\theta_{b}\over 2}-{i\pi\over 4}\right)
\cosh\left( {\theta-\theta_{b}\over 2}-{i\pi\over 4}\right)
\cosh \left({\theta+\theta_a\over 2}+{i\pi\over 4}\right)
\cosh \left({\theta-\theta_a\over 2}+{i\pi\over 4}\right)}\cr
=&{-2\over  \sinh\theta}\left(1-{R_b\over R_a}\right),\cr}}
together with
\eqn\idente{f(\theta+i\pi|\theta_{b},\theta_a)=
f^*(\theta|\theta_a,\theta_{b}),}
\eqn\addiident{f(\theta+2i\pi|\theta_{b},\theta_a)
=R_aR_b^*(\theta)f(\theta|\theta_{b},\theta_a),}
and
\eqn\moregood{f(\theta|
\theta_{b},\theta_a)={\kappa_b-i\sinh\theta\over
\kappa_b+i\sinh\theta}
f(-\theta|\theta_{b},\theta_a)
}

Based on these identities, we
 can now write the minimal solution for the form factor axioms
\axiom\ and \residucondi,
\eqn\FFmas{\eqalign{G(\theta_1,\theta_2)=&{i\over
8}\prod_{i=1,2}f(\theta_i)\ \sinh\theta_i\
{\kappa_b+i\sinh\theta_i\over
\kappa_b-i\sinh\theta_i}\ {1\over
\cosh({\theta_i\over 2}+i{\pi\over 4})}\cr &
\times \tanh\left({\theta_1-\theta_2\over 2}\right)
\tanh\left({\theta_1+\theta_2\over 2}\right),
}}%
{}From the expression of $G$ together with \identb, one easily
checks that
\eqn\monod{G(\theta_1+2i\pi,\theta_2|\theta_{b},\theta_a)=
-R_a(\theta_1)
R_b^*(\theta_1)
G(\theta_1,\theta_2|\theta_{b},\theta_a),}
so the residue axiom \residucondi\ is also satisfied.
The other form-factor $F$ can also be easily obtained.
For this, observe that
$$
F(\theta_1,\theta_2|\theta_{a},\theta_b)=\left[G(\theta_1,\theta_2|
\theta_b,\theta_{a})
\right]^*
$$
Using \identc,\identd,  the crossing identity follows:
\eqn\crossing{F(\theta_1,\theta_2)=-G(\theta_1-i\pi,\theta_2-i\pi),}
that is, explicitely
$$
_b\langle
0|\theta_1,\theta_2\rangle_a=_b\langle
\theta_1-i\pi,\theta_2-i\pi|0\rangle_a.
$$
It follows for instance that
\eqn\monodr{F(\theta_1+2i\pi,\theta_2|\theta_b,\theta_a)
=-R_a^*(\theta_1)R_b(\theta_1)G(
\theta_1,\theta_2|\theta_b,\theta_a).}

Finally, from \crossing, we deduce the result
\eqn\lastcross{{{}_b\langle\theta_2|\theta_1\rangle_a
\over {}_b\langle 0|0\rangle_a}=H(\theta_1,\theta_2)
=G(\theta_1-i\pi,\theta_2).}
It follows   that $H$ has a pole as $\theta_2\to
\theta_1$, with residue
$-i\left[1-R_a R_b^*(\theta_1)\right]$. It also has
a pole as $\theta_2\to -\theta_1$ with residue
$i\left[R_b(\theta_1)-R_a(\theta_1)\right]$.  As in the bulk case,
these poles
indicate  that \lastcross\ holds only for
``distinct'' rapidities \Smir;
the term
\eqn\delttt{
\delta(\theta_1-\theta_2)\left[1+R_aR_b^*(\theta_1)\right]+
\delta(\theta_1+\theta_2)\left[R_a(\theta_1)
+R_b(\theta_1)\right],}
has to be added at coincident rapidities.
This quantity, $H$, is probably the easiest to understand
intuitively. For instance, compare our
results with elementary computations for bare particles. With
boundary
condition $a$ at $x=0$ (and a trivial boundary
condition at $x=-L$),
a state $|\theta_a\rangle$ is associated a one particle wave function
$$
\psi_a(x,\theta)={1\over \sqrt{2}}\left(e^{im\sinh\theta
x}+R_a(\theta)
e^{-im\sinh\theta x}\right).
$$
The quantization condition is $e^{2im\sinh\theta L}R_a(\theta)=1$,
and the normalization
$$
\int_{-L}^0 |\psi|^2dx=L-{i\over m\sinh\theta}\hbox{Im
}R_a(\theta)\ \approx L, L>>1.
$$
One then finds the scalar product
$$
\eqalign{_b\langle\theta_2|\theta_1\rangle_a&=\int_{-L}^0
\psi_b^*(\theta_2,x)
\psi_a(\theta_1,x)\cr
&={i\over 2m}\left[{R_b^*(\theta_2)R_a(\theta_1)-1\over
\sinh\theta_2-\sinh\theta_1}+{R_b^*(\theta_2)-R_a(\theta_1)\over
\sinh\theta_2+\sinh\theta_1}\right],\cr}
$$
where we used the quantization conditions. This expression
has the same poles, with the same residues, as in \lastcross. The
complication
for the physical theory is  of course due to the fact that
the ground state shifts with different boundary conditions,
hence ``dressing'' the previous expression.

Inspection shows that $G$ has a simple pole at
$\theta_i={i\pi\over 2}$
and at $\theta_i=\pm\theta_b+{i\pi\over 2}$ (the pole of $\Phi$
at $\theta=\theta_B-{i\pi\over 2}$ is cancelled by the
$\kappa_b-i\sinh\theta$ term in $G$). Similarly, $F$ has a simple
pole at $\theta_i=-{i\pi\over 2}$ and at
$\theta_i=\pm\theta_a-{i\pi\over 2}$.
The poles of $G$ are thus the same as the poles of $R_b^*$, the poles
of $F$ the same as those of $R_a$.

Notice that the residue condition does not prevent the quantities
$F,G$
from having an added part without kinematic pole, with an expression
similar
to the ones written so far, but with the $\tanh$ of the sum and
difference
of rapidities replaced by the $\sinh$ function. However,
dimensional analysis, together with the requirement of having
a well defined massless limit and comparison with perturbation
theory,
exclude such a term.

Notice also that, like the $R$ matrices, the scalar products depend
only on squares of magnetic fields. It is easy to check that this
is true to the first  orders in perturbation theory, see the
appendix.
It is important to stress that our results apply in fact to the case
of
fields of the same sign; when $h_ah_b<0$, slightly different formulas
have to be used, which we discuss below.

We can  finally  compare some of these results with perturbation
theory. Consider the situation
where the two boundary fields are very small; at leading order,
the $R$ matrices and the functions $\Phi$ are evaluated for vanishing
fields, ie $\theta_{a}=\theta_b=i\pi$. One has in particular
$$
\Phi(\theta|i\pi,i\pi)=-{1\over \cosh\left({\theta\over 2}+i{\pi\over
4}\right)}.
$$
Putting these values
back in the formula, we find, after some algebra
$$
\eqalign{G(\theta_1,\theta_2)\propto &\tanh{\theta_1-\theta_2\over 2}
\tanh{\theta_1+\theta_2\over 2}\cosh\left({\theta_1\over
2}+i{\pi\over 4}\right)
\cr
&\times\cosh\left({\theta_2\over 2}+i{\pi\over
4}\right){\sinh\theta_1\sinh\theta_2\over
\cosh^2\theta_1\cosh^2\theta_2}.\cr}
$$
On the other hand, the scalar product
 can be computed from straightforward perturbation
theory.
The fermion propagators are obtained by solving the equations
of motion, using formula $(4.3)$ of \GZ, which provide directly
integral representations. After some lengthy algebra, one finds
a result in agreement with the previous formula for $G$.

\subsec{The general case}

Following the algebraic approach of \miwa\ we see that we could
as well put different boundary conditions in the left vacuum
to that of the right vacuum in the form factors
\eqn\ffvacch{
{\cal F}_{ba}(\theta_1,\cdots , \theta_n)={}_b \langle 0 |
{\cal O} Z^*(\theta_1)\cdots Z^*(\theta_n)
|0\rangle_{a}
}
Then the residue axiom for the two particle form factor
becomes
\eqn\newres{
{\rm Res} {\cal
F}_{ba}(\theta_1,\theta_2)\vert_{\theta_2=\theta_1+i\pi}=i
[1-R_a(\theta_1)R_b^*(\theta_1)] {}_b\langle O|\rangle_a
}
and can be generalised to the $n$ particles form factor in
the obvious fashion.
Here it is important to emphasize that the ${\cal F}$'s are not
form factors in a theory with inhomogeneous boundary
conditions, but rather  ``transition  form factors'' between
two theories having different boundary conditions (and thus
different Hilbert spaces) .
These axioms are general for operators which are local with
respect to the Fadeev-Zamolodchikov operators.  In particular,
if we choose
the case ${\cal O}=1$, ie the identity, then
the previous axioms will yield the scalar products (or transition
factors) between
asymptotic states  in Hilbert spaces with
different boundary conditions. The other axioms have to be modified
in a similar way; for instance, one has
\eqn\newji{
{\cal
F}_{ba}(\theta_1+2i\pi,\theta_2)=S(\theta_2-\theta_1)R_a^*(\theta_1)
S(-\theta_1-\theta_2)R_b^*(\theta_1+i\pi)
{\cal F}_{ba}(\theta_1,\theta_2).}
The modification of the exchange type relations is trivial.

In the special case of the Ising model  where $S=-1$, we get the
simpler equations (for a local operator)
\eqn\axiomising{\eqalign{
{\cal F}_{ba}(\cdots \theta_i+2 i
\pi\cdots)&=-R_a^*(\theta_i)R_b(\theta_i)
{\cal F}_{ba}(\cdots \theta_i\cdots) \cr
{\rm Res}_{\theta_j=\theta_1+i\pi}
{\cal F}_{ba}(\theta_1,\cdots ,\theta_i,\cdots \theta_n)&=
i(-1)^{j+1}\left[ 1-R_a(\theta_1)R_b^*(\theta_1)\right]
\cr&\times{\cal F}_{ba}(\hat{\theta_1},\cdots
\hat{\theta_j},\cdots,\theta_n)
\cr
{\cal F}_{ba}(\cdots \theta_i\cdots)&=R_a(\theta_i)
{\cal F}_{ba}(\cdots -\theta_i
\cdots). }}

Given the function $f$ constructed before, it is now easy to write
the generalisation of \FFmas:
\eqn\genform{\eqalign{
{{}_b\langle \theta_{2n},\cdots \theta_1|0\rangle_a
\over {}_b\langle 0|0\rangle_a}=&c_n \prod_i \sinh
\theta_if(\theta_i){\kappa_b+i\sinh\theta_i\over
\kappa_b-i\sinh\theta_i}
{1\over \cosh\left({\theta_i\over 2}+{i\pi\over 4}\right)}\cr
& \times\prod_{i< j} \tanh{\theta_i-\theta_j\over 2}
\tanh{\theta_i+\theta_j\over 2}
}}
with $c_n$ a proper normalisation determined by the residue
condition, $c_n=\left({i\over 8}\right)^n$.
The various other components are obtained by
crossing, in particular
\eqn\genformcrossed{\eqalign{
{{}_b\langle 0| \theta_{1},\cdots \theta_{2n}\rangle_a
\over  {}_b\langle 0|0\rangle_a}=&(-1)^nc_n \prod_i
 \sinh \theta_if(\theta_i-i\pi){\kappa_b-i\sinh\theta_i\over
\kappa_b+i\sinh\theta_i}
{1\over \cosh\left({\theta_i\over 2}-{i\pi\over 4}\right)}\cr
&\times \prod_{i<j} \tanh{\theta_i-\theta_j\over 2}
\tanh{\theta_i+\theta_j\over 2}
}}
These formulas also agree with the first non trivial order in
perturbation theory.

\subsec{Compatibility of form-factors and transition factors}

It must be clear that although asymptotic states with different
boundary conditions have some non trivial scalar products,
the form-factors of physical
operators have the same general form with any boundary
conditions, ie all
they see from the asymptotic states is their property
$||\theta>=R(\theta)||-\theta>$.
This is made possible by the axioms satisfied by the transition
factors, in particular
the pole condition. Let us illustrate this in the case of the
fermion operator. We have
\eqn\complki{{}_a\langle
0|\psi|\theta\rangle_a=\omega\left(e^{\theta/2}+R_a(\theta)
e^{-\theta/2}\right).}
On the other hand, we can compute this form-factor by inserting a
complete set
of states with $b$ boundary conditions on the left and on the right.
Using the behaviour
of transition factors under reflections, this allows us to reexpress
\complki\ as a sum of two
types of terms. The first type is
\eqn\complkii{\eqalign{&\omega\int {d\theta_1\over 2\pi}
e^{\theta_1/2}\left[
{}_a\langle 0|0\rangle_b{}_b\langle\theta_1|\theta\rangle_a\right.\cr
&\left. +\int {d\theta_2d\theta_3\over (4\pi)^22!}
{}_a\langle
0|\theta_2\theta_3\rangle_b{}_b\langle\theta_3\theta_2\theta_1|
\theta\rangle_a+\ldots\right],\cr}}
and the second type is
\eqn\complkiii{\eqalign{&\bar{\omega}\int {d\theta_1\over 2\pi}
e^{\theta_1/2}\left[
\int {d\theta_2\over 4\pi}{}_a\langle 0|\theta_1\theta_2\rangle_b
{}_b\langle\theta_2|\theta\rangle_a\right.\cr
&\left. +\int {d\theta_2d\theta_3d\theta_4\over (4\pi)^3 3!}
{}_a\langle 0|\theta_1\theta_2\theta_3\theta_4\rangle_b{}_b\langle
\theta_4\theta_3\theta_2|
\theta\rangle_a+\ldots\right].\cr}}
In all these formulas integrals are regulated by taking principal
parts. Let us now consider the
second sum \complkiii. The $\theta_1$ integral has no singularity on
the real axis;
we move the contour to $\hbox{Im }(\theta_1)=\pi$; by doing so we do
not encounter
any singularity since the transition factors with $\theta_1$ on the
right have
poles only in the lower half plane. By using crossing and the
completude
 relation with $b$ boundary condition,
this gives rise to $\omega\left(e^{\theta/2}+R_a(\theta)
e^{-\theta/2}\right)$, up to two corrections. First,  the correct
crossing
formula is ${}_a\langle
0|\theta_1+i\pi,\theta_2\rangle_b={}_a\langle\theta_1|
\theta_2\rangle_b^{dc}$, where on the right we mean the disconnected
transition factor,
without the added delta function part at coincident rapidities. On
the other
hand, the completude relation works with ${}_a\langle\theta_1|
\theta_2\rangle_b$: this means that we have to subtract a first
correction,
which reads using \delttt, together with a bit of algebra,
like \complkii\ but with the integrand for the
$\theta_1$ variable multiplied by $1+{1\over 2}{R_a\over
R_b}(\theta_1)
+{1\over 2}{R_b\over R_a}(\theta_1)$.
Second, after crossing, we get a $\theta_1$ integral that runs over
the whole real
axis, avoiding the poles at $\theta_2$ and $-\theta_2$ by going under
them. The
completude relation on the other hand requires the principal part
integrals, so
we have to subtract the second correction, given by the pole
contributions. This gives
rise again to a formula like \complkii\ but with the integrand for
the $\theta_1$
variable multiplied by $1-{1\over 2}{R_a\over R_b}(\theta_1)
-{1\over 2}{R_b\over R_a}(\theta_1)$, where we used the residue
conditions for $H$. Adding up the two corrections, we see that the
second type of terms \complkiii\ is nothing but \complki\ minus the
first
 type of terms \complkii.
Hence, adding up \complkii\ and \complkiii\ gives \complki, and we
recover
our form factor.

\newsec{Massless limit and boundary conditions changing operators}

We now consider the case where the bulk is in fact massless. This
is conveniently adressed within the framework of massless
scattering \ref\FS{P. Fendley, H.Saleur, in Proceedings of the
Trieste Summer School on high Energy Physics and Cosmology, July
1993,
World Scientific, singapore (1993).},
\ref\ZZ{A.B. Zamolodchikov, Al.B. Zamolodchikov, Nucl. Phys.
{\bf B}379 (1992), 602.}, which  is formally
obtained by letting
the physical mass $m\to 0$, together with the rapidities
$\theta\to\pm\infty$.
We set $m={\mu\over 2}e^{-\Theta}$ ($\mu$ an arbitrary scale, taken
equal to unity in what follows),
 $\theta=\pm \Theta\pm\beta$,
 to parametrize energy and momentum $e=\pm p=\mu e^\beta$. The $\pm$
sign
corresponds to R (L) movers.

We first consider time propagation in the x-direction (direct
channel). Then, L and R
movers are defined by the conventions, eg for the form-factors of the
fermions
\eqn\defi{\eqalign{\langle 0|\psi_R(x,y)|\beta\rangle_R&=\omega
e^{\beta/2}
\exp\left[e^\beta (x+iy)\right]\cr
\langle 0|\psi_L(x,y)|\beta\rangle_L&=\bar{\omega}
e^{\beta/2}\exp\left[e^\beta(x-iy)\right].\cr}}
Consider then the one point function of the energy operator,
$\epsilon\equiv
i\psi_L\psi_R$. Using
the expression for the boundary state
\eqn\bdrstate{
|B\rangle =\sum_{n=0}^\infty {1\over n!}\int
\prod_i {d\beta_i\over 2\pi}
K(\beta_i-\beta_B)
Z_L^*(\beta_i)Z^*_R(\beta_i)|0\rangle,}
one finds (here, $z=x+iy$)
\eqn\enonept{\langle \epsilon(z,\bar{z})\rangle=\langle 0|
\epsilon(z,\bar{z})|B\rangle=-i
\int {d\beta\over 2\pi} K(\beta-\beta_B)
e^\beta e^{2 e^\beta x}.}
In the case of fixed boundary conditions $K=i$, and one finds
$\langle \epsilon\rangle=-{1\over 4\pi x}$. For free boundary
conditions,
$K=-i$, and one finds the opposite result.

In contrast with the massive case, where the expression \enonept\
could
not be obtained directly   in the cross-channel, such a computation
is now possible.
This is because we can represent
asymptotic states as superpositions of left and right moving
parts,
the simplest ones being\foot{These states are normalized such that,
for
instance, $\langle
\beta_1||\beta_2\rangle=2\pi\delta(\beta_1-\beta_2)$}
\eqn\asympst{\eqalign{||\beta\rangle=&|\beta\rangle_R+
R(\beta)|\beta\rangle_L\cr
||\beta_1\beta_2\rangle=&|\beta_1,\beta_2\rangle_{RR}+R(\beta_2)
|\beta_1\beta_2\rangle_{RL}
+R(\beta_1)|\beta_1\beta_2\rangle_{LR}\cr&+R(\beta_1)R(\beta_2)
|\beta_1\beta_2\rangle_{LL},\cr}}
and use the LR factorized form of conformal operators. Here, the
meaning of L and R is encoded in
the new dependence of form-factors
\eqn\newdep{\eqalign{\langle 0||\psi_R(x,y)||\beta\rangle_R&
=\omega e^{\beta/2} \exp\left[e^\beta
(-y+ix)\right]\cr
\langle 0||\psi_L(x,y)||\beta\rangle_L&=\bar{\omega}
e^{\beta/2}\exp\left[e^\beta(-y-ix)\right].\cr}}
Using the general crossing formula
written earlier,
the one point function of the energy reads then ($w=iz$)
\eqn\crosch{\langle \epsilon(w,\bar{w})\rangle =-i\int
{d\beta\over 2\pi}
\langle 0||\psi_L(\bar{w})||\beta\rangle
\langle\beta||\psi_R(w)||0\rangle =\int
{d\beta\over 2\pi}
R(\beta-\beta_B)e^\beta e^{-2ixe^\beta}.}
A rotation of the countour shows that the expressions \enonept\
and \crosch\
are identical, with the identification $K(\beta)=R({i\pi\over
2}-\beta)$, as in the massive case. The massless reflection matrices
actually follow from the massive ones  by setting
 $\theta_b\approx \Theta+\beta_b$: they are of the form
 $R=i\tanh\left({\beta-\beta_b\over 2}-{i\pi\over 4}\right)$.

We will define massless form-factors as form-factors on the previous
asymptotic states.
They follow from the massive case by taking the infinite rapidity
limit $\theta=\pm\Theta\pm\beta$
for each rapidity. For instance
\eqn\twopartff{\eqalign{\langle 0||\epsilon(w,\bar{w})
||\beta_1\beta_2\rangle
=i
e^{\beta_1/2}e^{\beta_2/2}\left\{
R(\beta_2)\exp[e^{\beta_1}(-y+ix)+e^{\beta_2}(-y-ix)]-\right.\cr
\left.R(\beta_1)\exp[e^{\beta_1}(-y-ix)+
e^{\beta_2}(-y+ix)]\right\}.\cr}}
The scalar products between asymptotic states with
different boundary conditions follow
by the same limiting procedure. Alternatively, one could obtain them
by writing a similar set of massless axioms.
As an example, in the massless case, the expressions for the
one point function of the energy  are
\eqn\expresi{\eqalign{\langle \epsilon(x,y)\rangle_{ba}=\int
{d\beta\over 2\pi}
R_b(\beta)e^\beta e^{-2ixe^\beta}
-i\int {d\beta_1d\beta_2\over 8\pi^2} R_b(\beta_1)\cr
G(\beta_1,\beta_2) e^{\beta_1/2}
e^{\beta_2/2}\exp\left[
e^{\beta_1}(-y-ix)+e^{\beta_2}(-y+ix)\right],\cr}}
for $y>0$,and
\eqn\expresi{\eqalign{\langle \epsilon(x,y)\rangle_{ba}=\int
{d\beta\over 2\pi}
R_a(\beta)e^\beta e^{-2ixe^\beta}
+i\int {d\beta_1d\beta_2\over 8\pi^2} R_a^*(\beta_1)\cr
G(\beta_1,\beta_2) e^{\beta_1/2}
e^{\beta_2/2}\exp\left[
e^{\beta_1}(y+ix)+e^{\beta_2}(y-ix)\right],\cr}}
for $y<0$ (where we used the antisymmetry of $F,G$)
leading to the  same condition as \residucond\ with now
the massless reflection matrices and in the
variable $\beta$.
Using that $f(\theta|\theta_b,\theta_a)\approx -2
e^{-{\Theta+\beta\over 2}}
\sqrt{\sinh{\beta_b-\beta_a\over 2}}\Phi(\beta|\beta_b,\beta_a)$
gives the result
\eqn\atlast{G(\beta_1\beta_2)=-{1\over 2}
\sinh{\beta_b-\beta_a\over
2}\left[\prod_{i=1,2}\Phi(\beta_i|\beta_b,\beta_a)
{e^{\beta_i}+iT_b\over
e^{\beta_i}-iT_b}\right]\tanh{\beta_1-\beta_2\over 2}}
where we have set $T_{a(b)}=e^{\beta_{a(b)}}$.

We now consider the case where the mixed boundary conditions
are conformal invariant: we chose free boundary conditions (F)
for $y<0$ and fixed boundary conditions (+) for $y>0$.
This is described in the previous
formalism by taking the limits $\beta_a\to -\infty$ and
$\beta_b\to\infty$.
{}From  previous formulas, we find
\eqn\postu{{_+\langle\beta_2\beta_1|0\rangle_F
\over _+\langle0|0\rangle_F}= i\tanh{\beta_1-\beta_2\over 2}.}
The one point function of the energy reads then
\eqn\newoneptft{\eqalign{
\langle\epsilon(w,\bar{w})\rangle_{+F}=&\ i\  \ \int
{d\beta_1d\beta_2\over
4\pi^2}
e^{\beta_1/2}e^{\beta_2/2} \tanh({\beta_1-\beta_2\over 2})
\exp[ix(e^{\beta_1}-
e^{\beta_2})]\cr
&\times \exp[-y(e^{\beta_1}+e^{\beta_2})] - {1\over 4\pi x}.\cr}}
Explicit evaluation gives
\eqn\newoneptfti{\langle\epsilon(w,\bar{w})\rangle_{+F}={1\over 4\pi}
\left({1\over x}
-{y\over x\sqrt{x^2+y^2}}\right)-{1\over 4\pi x},}
where we used the formula
\eqn\form{\int_0^\infty {ds_1\over \sqrt{s_1}}{ds_2\over \sqrt{s_2}}
i{s_1-s_2\over s_1+s_2}e^{ix(s_1-s_2)}e^{-y(s_1+s_2)}=\pi\left(
{1\over x}-{y\over x\sqrt{x^2+y^2}}\right).}
On the other hand, one easily gets from conformal field theory
results
\Burk\
that
$$
\langle\epsilon(w,\bar{w})\rangle_{+F}=-{1\over 4\pi x}{y\over
\sqrt{x^2+y^2}},
$$
in agreement with \newoneptfti.

By duality, one gets similarly
\eqn\postui{{_F\langle\beta_2\beta_1|0\rangle_+\over
_F\langle0|0\rangle_+}=i
\tanh{\beta_1-\beta_2\over 2}
.}
The  computation leading to \newoneptfti\ was done for $y>0$. By
considering
$y<0$
instead, one also gets
\eqn\postuii{{_+\langle0|\beta_1\beta_2\rangle_F\over
_+\langle0|0\rangle_F}=-i
\tanh{\beta_1-\beta_2\over 2}.}

In a similar way, we can consider the two point function of fermions.
By considering the cases where $y_1y_2>0$ and $y_1y_2<0$,
one also
finds
the result
\eqn\crossedff{{\ _+\langle\beta_2|\beta_1\rangle_F\over \
_+\langle0|0\rangle_F}=i
\coth{\beta_1-\beta_2\over 2}.}

The previous formula are compatible with the
standard description \JCardy\
of mixed  conformal invariant boundary conditions,
which involves putting a ``boundary
conditions
changing operator'' at $y=0$. In the case
from free to fixed spins, this operator has to be the $\phi_{12}$
operator of conformal weight ${1\over 16}$.
{}From this identification, it follows  that
 \postu\ generalizes to
\eqn\morepostu{ {_+\langle\beta_{2n}\ldots \beta_{1}|0\rangle_F
\over _+\langle0|0\rangle_F}=   i^n \prod_{i<j}\
\tanh{\beta_i-\beta_j\over 2},}
and that the scalar products coincide, formally,
 with the form factors of the  (eg right moving) spin operator in
the bulk. This ensures in particular that the partition fucntion of
an Ising model with a region of fixed boundary conditions of length
$L$
inserted into a wall of free boundary conditions decays as
$L^{-1/8}$.

\newsec{The case $h_ah_b<0$}

As mentioned earlier,  we expect the results of section 3  to hold
for the case
 $h_ah_b>0$ only; physically, this corresponds to a situation
where
 the boundary
condition at $x=-\infty$ is independent of $y$. Instead, when
$h_ah_b<0$
 and
large enough magnetic fields, spins for $y<0$ and $y>0$
tend to
be oriented in opposite directions, in particular they go to $\pm
{\cal M}$,
 where ${\cal M}$ is the spontaneous magnetization, at $x=-\infty$.
There is
thus a frustration line inserted in the system - in other words,
following standard mappings \ref\KC{L.P. Kadanoff, H. Ceva,
Phys. Rev. B3  (1971), 3918.},
a spin operator
is inserted at $x=y=0$. Of course, the spin operator on the boundary
acquires
 a dimension $1/2$, so in fact we have   a fermion line emitted at
the transition.
We thus expect that the problem of $+-$ boundary conditions will be
described by
transition factors involving an odd number of particles; from
 that perspective,
the ground state itself must be thought of as containing a particle
of vanishing energy, and momentum equal to $im$ (so the partition
function
of a model with a frustration line extending to $x$ goes as $e^{mx}$)
, ie a particle at
 rapidity $\theta={i\pi\over 2}$,  which we denote by $I$ in what
follows.

It is easy to check that this description is adequate
in the massless case, and for boundary conditions $a=+$, $b=-$,
ie the transition from ``fixed up'' to ``fixed down''.
  From conformal invariance, one expects
\Burk,

\eqn\newstuff{\langle \epsilon(w,\bar{w})\rangle_{+-}=
\langle \epsilon(w,\bar{w})\rangle_{-+}=-{1\over 4\pi
x}\left(1-4{x^2\over x^2+y^2}\right).}
Clearly, this cannot follow from formulas in section 3,
since our scalar products $F$ and $G$, which are functions
only of squares of fields,  vanish for $h_a=\pm h_b$. On the other
hand,
formula \newstuff\ follows from the description with an odd number of
particles.
To see this, observe first
that in the massless case, the particle at imaginary rapidity has
vanishing energy and momentum,
and  merely  keeps the correct parity of fermion numbers.
Consider then the one point function of the energy. Two processes can
contribute to it:
in one case, the transition simply emits the particle at imaginary
rapidity, that just
``goes through'' the energy insertion without being affected, so the
process
contributes $\langle\epsilon\rangle_+=-{1\over 4\pi x}$. The other
process occurs when the
transition emits a particle at real rapidity, which is then
transformed into
 the particle at imaginary rapidity by the energy insertion.
The required energy form factor is obtained
by taking the usual limit:
$$
\langle 0||\epsilon(w,\bar{w})||\beta,I\rangle=\omega e^{\beta/2}
\left\{\exp\left[e^\beta(-y+ix)\right]-
\exp\left[e^\beta(-y-ix)\right]
\right\}
$$
As for the scalar product ${}_+\langle\beta|0\rangle_-$,  it should
be of the form $\lambda e^{\beta/2}$ ($\lambda$ an unknown constant)
since
 the frustration line emitted by the boundary corresponds
to a fermion insertion, of dimension $1/2$ (in other words,
the partition function of an Ising model with
a region of $-$ boundary conditions of length $L$
inside a domain of $+$ boundary conditions decays as $\int d\beta
e^\beta
\exp\left[-e^{\beta}L\right]= {1\over L}$, as
required from conformal invariance considerations \JCardy).
 From this, we get the second contribution
to the energy one point function
$$
\lambda\omega\int {d\beta\over 2\pi} e^{\beta}
\left\{\exp\left[e^\beta(-y+ix)\right]-
\exp\left[e^\beta(-y-ix)\right]
\right\}
={i\lambda\omega\over\pi} {x\over x^2+y^2},
$$
thus recovering \newstuff\ provided $\lambda=-\omega$.

As a further check, we now compute the two point function of the
energy with
the $+-$ boundary conditions. Conformal invariance gives the result
\eqn\evennewer{\eqalign{\langle \epsilon(1)\epsilon(2)\rangle_{+-}=
{1\over 4\pi^2}\left\{{1\over 4 x_1x_2}\left(1-4\sin^2\alpha_1\right)
\left(1-4\sin^2\alpha_2\right)\right.\cr
\left. +{1\over
r^2}\left[1-4\sin^2\left(\alpha_1-\alpha_2\right)\right]
-{1\over
r^2+4x_1x_2}\left[1-4\sin^2\left(\alpha_1+\alpha_2\right)\right]
\right\},\cr}}
where $r^2=\left(x_1-x_2\right)^2+\left(y_1-y_2\right)^2$,
$\sin\alpha={x\over \sqrt{x^2+y^2}}$.
Several processes are now allowed, without or with
real particles  emitted from the transition. In the latter case,
processes
 where the
real particles emitted at the transition are destroyed by only one of
the energy
operators are identical to processes appearing in the evaluation of
\newstuff.
The only new processes, which we call interacting processes, must add
up to,
using \evennewer,
$$
{1\over 4\pi^2}\left[{4\sin^2\alpha_1\sin^2\alpha_2\over x_1x_2}
-{4\over r^2}\sin^2\left(\alpha_1-\alpha_2\right)+{4\over
r^2+4x_1x_2}
\sin^2\left(\alpha_1+\alpha_2\right)\right]
$$
After a bit of algebra, this reads
\eqn\stepi{
{1\over r^2+4x_1x_2}{
x_1x_2\left(r_1^4+r_2^4\right)
 -2x_1x_2y_1^2y_2^2-2x_1^3x_2^3-2x_1x_2\left(x_1^2y_2^2+
x_2^2y_1^2\right)\over\pi^2 r_1^2r_2^2r^2},}
where $r_1^2=x_1^2+y_1^2$. On the other hand, there are two
interaction processes. In the first one,
the transition can emit a particle at rapidity $\beta_1$, the first
insertion
 of the energy can destroy the
particle $\beta_1$ and create one at rapidity $\beta_2$, while
 the second insertion of the energy destroys the latter to replace it
by the
 imaginary one. This process has the amplitude
$$
\eqalign{-\int {d\beta_1\over 2\pi}{d\beta_2\over 2\pi}
e^{\beta_1}e^{\beta_2}\left\{\exp\left[e^{\beta_1}
\left(-y_1-ix_1\right)
-e^{\beta_2}\left(-y_1+ix_1\right)\right]\right.\cr
\left. - \exp\left[e^{\beta_1}\left(-y_1+ix_1\right)
-e^{\beta_2}\left(-y_1-ix_1\right)\right]\right\}\cr
\left\{
\exp\left[e^{\beta_2}\left(-y_2-ix_2\right)\right]-
\exp\left[e^{\beta_2}\left(-y_2+ix_2\right)\right]\right\}.\cr}
$$
After a bit of algebra, this reads
\eqn\stepii{
{1\over 4\pi^2}\left({2\over r^2+4x_1x_2}{y_1y_2-x_1x_2-r_1^2\over
r_1^2}
-{2\over r^2}{y_1y_2+x_1x_2-r^2_1\over r_1^2}\right).}
In the second process, the transition emits a particle at rapidity
$\beta_1$,
the first
energy insertion adds up a particle at rapidity $\beta_2$ and
the particle at imaginary rapidity (the particle at rapidity
$\beta_1$
just going through), and the two real particles
are finally destroyed by the second energy insertion. This process
has the amplitude
$$
\eqalign{-\int {d\beta_1\over 2\pi}{d\beta_2\over 2\pi}
e^{\beta_1}e^{\beta_2}\left\{\exp\left[e^{\beta_1}
\left(-y_2-ix_2\right)
+e^{\beta_2}\left(-y_2+ix_2\right)\right]\right.\cr
\left. - \exp\left[e^{\beta_1}\left(-y_2+ix_2\right)
+e^{\beta_2}\left(-y_1-ix_2\right)\right]\right\}\cr
\left\{
\exp\left[-e^{\beta_2}\left(-y_1-ix_1\right)\right]-
\exp\left[-e^{\beta_2}\left(-y_1+ix_1\right)\right]\right\}.\cr}
$$
which  is the same as the previous result, with $1$ and $2$
interchanged.
Adding \stepii\ and the similar result with $1$ and $2$ interchanged
reproduces  \stepi\ indeed.

We can now complete our study of scalar products by giving results
equivalent to those
of section 3 when $h_ah_b<0$. Now only odd numbers of particles
are emitted. The amplitude for an even number of real particles is
\eqn\genformtata{
{{}_b\langle \theta_{2n}\ldots \theta_1, I|0\rangle_a
\over {}_b\langle I|0\rangle_a}=G(\theta_1,\ldots,\theta_{2n}),}
where the expression on the right hand side coincides with \genform,
while
for an odd number of real particles we have the obvious
generalization
of \genform
\eqn\genformi{\eqalign{
{{}_b\langle \theta_{2n+1},\cdots \theta_1|0\rangle_a
\over {}_b\langle I|0\rangle_a}&=c_n \prod_i
\sinh \theta_if(\theta_i){\kappa_b+i\sinh\theta_i\over
\kappa_b-i\sinh\theta_i}
{1\over \cosh\left({\theta_i\over 2}+{i\pi\over 4}\right)}\cr
& \prod_{i<j} \tanh{\theta_i-\theta_j\over 2}
\tanh{\theta_i+\theta_j\over 2}\cr
}}
The residue axioms fix $c_n$ again, except for an overall scale this
time:
$c_n=c \left({i\over 8}\right)^n$.

To fix this scale, we observe that at least the one particle
form-factor
should not vanish when $h_a=-h_b$, and that it should vanish when
$h_ah_b=0$,
since then there is no frustration left. Recall that
 $f$ itself goes as $\sqrt{h_b^2-h_a^2}$;
we thus chose $c_n$ to be proportional to
$\sqrt{h_ah_b}{h_a-h_b\over\sqrt
{h_b^2-h_a^2}}$, ie
we expect the one particle scalar product
\eqn\cute{{}_b\langle\theta|0\rangle_a={\omega\over 2\sqrt{2}}
\sqrt{h_ah_b}{h_a-h_b\over\sqrt{h_b^2-h_a^2}}
\sinh\theta f(\theta)
{\kappa_b+i\sinh\theta\over \kappa_b-i\sinh\theta}
{1\over \cosh\left({\theta\over 2}+{i\pi\over 4}\right)},}
where the field dependent prefactor and numerical factors are matched
to the
massless limit. Indeed, in that  limit, $h\approx e^{\beta/2}$,
giving rise to
$$
{}_b\langle\beta|0\rangle_a=-{1\over 2}
\left(h_a-h_b\right)\Phi(\beta|\beta_b,\beta_a){e^\beta+iT_b\over
e^\beta-iT_b}.
$$
Finally, in the limit of large fields $h_a,h_b\to\infty$, $h_a/h_b$
finite,
this goes, using $\Phi(\beta|\beta_b,\beta_a)\approx 2e^{i\pi/4}
e^{\beta/2} e^{-\beta_a/4}e^{-\beta_b/4}$, to
$$
{}_b\langle\beta|0\rangle_a=\omega{h_a-h_b\over\sqrt{h_ah_b}}
e^{\beta/2}
$$
In the case of opposite fields we recover the previous result for
 $+-$ boundary conditions.

Observe that, when $h_a=-h_b$, only the one particle scalar product
is non zero,
due to the terms $\sqrt{\kappa_a-\kappa_b}$ in the functions $f$ (
since the only undetermined quantity is an overall normalization, we
cannot change
this without spoiling the results in the one particle case).

In the perturbative limit, when $h_a$ and $h_b\to 0$, one has
\eqn\morepertodd{
{}_b\langle\theta|0\rangle_a\propto\sqrt{h_ah_b}
\left(h_a-h_b\right){\sinh\theta\over\cosh\theta\cosh
\left({\theta\over 2}
-{i\pi\over 4}\right)}.}
This can be compared with perturbative results. To do so, we have
to slightly modify the action \act\ to take into account  the fermion
line
 emitted at the origin. It is natural to represent it by a term
$\sqrt{h_ah_b}\psi_0a(0)$,
where $a$ is the same boundary fermionic degree of freedom as in
\act,
the term $\sqrt{h_ah_b}$ is dictated by $a,b$ symmetry and
dimensional
analysis, and $\psi_0$ is a fermion zero mode obeying
$\langle\psi_0\psi_{L,R}\rangle
\propto e^{mx}$. There are now two  non trivial scalar products
at  second order: one does not involve insertion of this term, and
thus
is identical
as the one we had for $h_ah_b>0$, in agreement with \genformtata. The
second involves insertion of this term, plus a single integral along
either the region $y<0$ or $y>0$. This goes as
$(h_a-h_b)\sqrt{h_ah_b}\int_{-\infty}^0
dy \langle \left(\psi_L+\psi_R\right)(x=0,y)\psi_L(x,y)\rangle$,
and the integral can be shown, using the propagators in the appendix,
 to agree with \morepertodd.

\newsec{Sinh-Gordon model.}

In this section we proceed to try to construct some of
the transition factors for the
sinh-Gordon model.  Our  purpose here is
to gain control of the axiomatic relations satisfied by these
factors in cases where the S matrix is non trivial, hence
paving the way for further study of the sine-Gordon case,
which is the most important for applications. As we will see,
some complications arise when we wish to identify operators
creating the need for a more systematic study.

Let us first introduce some notation. The
action is
\eqn\shaction{
A=\int_{-\infty}^0 dx \int_{-\infty}^\infty dy
\ \left[
{1\over 2} (\partial_\mu \phi)^2-{m_0^2\over b^2}
\cosh b\phi\right] +
\lambda \int_{-\infty}^\infty dy \cosh {b\over
2}(\phi-\phi_0)(x=0,y).
}
The two particle $S$ matrix and the
reflection matrix for this model are well
known. Setting
\eqn\noti{\eqalign{
\xi (a)&={\sinh({\theta\over 2}+i{\pi a\over 4})\over
\sinh({\theta\over 2}-i{\pi a\over 4})} \cr
&=\exp\left[ -4\int_0^\infty {dt\over t} {\sinh{a t\over 4}
\cosh(1-{a\over 4})t
 \over \sinh t} \sinh({\theta t\over i \pi})\right]  , \ \
0<{\rm Re} \ a< 4,
}}
they are given by
\eqn\srmat{\eqalign{
S&=-{1\over \xi (B) \xi(2-B)}\cr
R&={\xi (1)\xi (2-B/2) \xi(1+B/2)
\over \xi (1-E(b))\xi (1+E(b))\xi (1-F(b))
\xi (1+F(b))},
}}
where $B={1\over 2\pi} {b^2\over 1+b^2/4\pi}$, and
$E,F$ are functions of $b$, also depending on the boundary
parameters: they are given
by $E={B \eta\over \pi}$ and $F=i {B \Theta\over \pi}$
in Ghoshal's notation \ref\Gh{S. Ghoshal, Int. J. Mod. Phys.
A{\bf 9} (1994), 4801.}.

In the following we will choose to
work with the boundary action \shaction\ having the phase $\phi_0=0$
which
leads to $\eta=E=0$ and simplifies the reflection matrix.

Associated with the product
$\xi (1+a) \xi(1-a)$ we define $\Upsilon (a)$ by
\eqn\gena{\eqalign{
\Upsilon (a,\theta)&={\cal N}_a \exp 2\left\{ \int_0^\infty
{dt\over t} {\cosh{a t\over 2}\over
\cosh {t\over 2} \sinh t} \sin^2({\hat{\theta} t\over 2\pi})\right\}
\cr
{\cal N}_a &=\sqrt{\cos {\pi a\over 2}}
\exp {1\over 2}\left[ \int_0^\infty
{dt\over t} {\cosh{at\over 2} \sinh {t\over 2}\over
\cosh^2{t\over 2}}\right],
}}
with $\hat{\theta}=i\pi-\theta$.
This function is constructed so that it satisfies
\eqn\obvious{
\Upsilon (a,\theta+2\pi i)=\Upsilon (a,-\theta)=
{1\over \xi (1+a)\xi(1-a)}
 \ \Upsilon (a,\theta)
}
and has no poles nor zeroes in the physical strip.
We will need this function in the next section to construct
the form factors.  Our choice of normalisation is such
that
\eqn\normups{
\Upsilon(a,\theta)\Upsilon(a,\theta+i\pi)=\sinh\theta+
i\cos{\pi a\over 2}.
}

Boundary form-factors as well as transition factors
can now be obtained by solving the set of axioms written in
the previous sections.  Let us start by investigating
some general properties.

\subsec{Two particles sector.}

The two particle sector  is where differences with the Ising model
 are emerging.
Let us suppose that  we have a set of boundary operators (local)
denoted generically by $k$, then
the transition form factor between states in theory $a$ and theory
$b$ can
be inferred to have the form
\eqn\ffabsi{
F_{ab}^k(\theta_1,\theta_2)=c_2^k \ f_{ab}(\theta_1)f_{ab}(\theta_2)
\times
\left\{ {F_{min}(\theta_1-\theta_2) F_{min}(\theta_1+
\theta_2)\over \cosh({\theta_1-\theta_2\over 2})
\cosh({\theta_1+\theta_2\over 2})
}\right\} Q^k(\theta_1,\theta_2).
}
In this formula, $F_{min}$ is the usual function generating
the sinh-Gordon $S$ matrix
\eqn\fminexp{\eqalign{
F_{min}(\theta)&={\cal N}_B \exp \left[8\int_0^\infty
{dt\over t} {\sinh({t B\over 4}) \sinh\left[{t\over 2}\left(1-{B\over
2}\right)\right]
\sinh{t\over 2}\over \sinh^2 t} \sin^2\left({t\hat{\theta}\over
2\pi}\right)
\right] \cr
{\cal N}_B&=\exp \left[-4\int_0^\infty
{dt\over t} {\sinh({t B\over 4}) \sinh\left[{t\over 2}\left(1-{B\over
2}\right)\right]
\sinh{t\over 2}\over \sinh^2 t}\right].
}}
All the dependence on the boundary is contained
in the factors $f_{ab}$.  $Q^k$ is a polynomial we choose to be
invariant under reflections, exchange and $2\pi i$ shifts.
This leads to the following conditions on the $f_{ab}$'s
\eqn\condfab{\eqalign{
f_{ab}(\theta)&=R_b(\theta) f_{ab}(-\theta)
\cr
f_{ab}(\theta+2\pi i)&=R_a^*(\theta+i\pi) R_b^*(\theta)
f_{ab}(\theta)
}}
for which a minimal solution can easily be found.  Using the crossing
\eqn\cross{
R\left(i{\pi\over 2}-\theta\right)=S(2\theta) R\left(i{\pi\over
2}+\theta\right)
}
we can write this last relation as
\eqn\ndeux{
f_{ab}(\theta+2\pi i)={R_a(\theta) R_b^*(\theta)\over S(2\theta)}
f_{ab}(\theta).
}
In order to construct this function, we start with the functions
$u$ and $v$ satisfying
\eqn\uvfunc{\eqalign{
u(\theta)u(\theta+i\pi)&={1\over F_{min}(2\theta+i\pi)}\cr
v(\theta+2\pi i)&={R_a(\theta)\over R_b(\theta)} v(\theta).
}}
The function $u$ will generate the $S(2\theta)$ and is easily
found to be:
\eqn\ufunc{
u(\theta)=\exp 2\left[\int_0^\infty {dt\over t}
{\sinh{B t\over 4} \sinh\left[\left(1-{B\over 2}\right){t\over
2}\right]
\sinh{t\over 2}\over
\sinh^2 t \cosh t} \cos\left[\left({2\theta\over \pi}-i\right)
t\right]
\right],
}
and $v$ is given by
\eqn\vfunc{
v(\theta)={\Upsilon (F_a,\theta) \over
\Upsilon (F_b,\theta)}.
}

The superscript refers to  boundary conditions $a,b$, corresponding
to   parameters $F_a,F_b$ (since we have set $\phi_0=0$ in the
action,
$a$ and $b$ correspond simply to different values of the coupling
$\lambda$).
Now that the monodromy relations are satisfied, we need to multiply
by other terms to get the correct reflection equation.
This leads to the minimal solution
\eqn\fabint{
f_{ab}(\theta)={\sinh\theta \over \sinh\left[{\theta\over 2}+i{\pi
(1-F_a)\over 4}\right] \sinh\left[{\theta\over 2}+i{\pi (1+F_a)\over
4}\right]
}
u(\theta) {\Upsilon (F_a,\theta)\over
\Upsilon (F_b,\theta)}
}
where we used the fact that $u(\theta)/u(-\theta)=-
{\xi (1+B/2)\xi (2-B/2) \over \xi (1)}$.
This minimal solution satisfies
\eqn\ufunc{\eqalign{
f_{ab}(\theta) f_{ab}(\theta+i\pi)&={-4\sinh^2\theta \over
\cosh(\theta-i{\pi F_a\over 2})\cosh(\theta+i{\pi F_a\over 2})}
\left({\sinh\theta+i\cos{\pi F_a\over 2}\over
\sinh\theta+i\cos {\pi F_b\over 2} } \right)
{1 \over F_{min}(2\theta+i\pi)}
\cr
&={-4\sinh^2\theta \over
(\sinh\theta-i\cos{\pi F_a\over 2})
(\sinh\theta+i\cos {\pi F_b\over 2} )}
{1 \over F_{min}(2\theta+i\pi)}
}}
This, added to the residue
axiom leads to an equation for the polynomial $Q^k$.
Before embarking on the explicit form of these polynomials
for specific operators, let us generalize these results to
the $N$ particle sector.

\subsec{N particle sector.}

Having understood the two particle sector the
general structure emerges nicely and turns out to be a
very simple generalisation of the two particle case:
we have
\eqn\ffabn{\eqalign{
F_{ab}^k(\theta_1,\cdots,\theta_n)&=c_n^k \
\prod_{i=1}^n f_{ab}(\theta_i) \times \cr &
\times
\prod_{i<j}
\left\{ {F_{min}(\theta_i-\theta_j) F_{min}(\theta_i+
\theta_j)\over \cosh({\theta_i-\theta_j\over 2})
\cosh({\theta_i+\theta_j\over 2})
}\right\} Q^k(\theta_1,\cdots, \theta_n).
}}
By construction, this satisfies all the exchange and reflection
relations.
Here only the residue equation needs to be fixed: it
leads to an equation for the polynomial $Q^k$ in
the form
\eqn\eqsympol{
Q^k(\theta+i\pi,\theta,\theta_3,\cdots,\theta_n)=P(\theta | \theta_3,
\cdots \theta_n)
Q^k(\theta_3,\cdots,\theta_n),
}
with $P$ given by
\eqn\ppoly{\eqalign{
P(\theta|\theta_3\cdots\theta_n)=&{1\over \sinh\theta}
\left\{\left(\sinh\theta-i\cos{\pi F_a\over 2}\right)
\left(\sinh\theta+i\cos{\pi F_b\over 2}\right)\times \right.
\cr &
\prod_{i=3}^n \left[\sinh(\theta-\theta_j)+\sinh{i\pi B\over
2}\right]
\left[\sinh(\theta+\theta_j)+\sinh{i\pi B\over 2}\right]
\cr &
-\left(\sinh\theta+i\cos{\pi F_a\over 2}\right)\left(\sinh\theta-
i\cos{\pi F_b\over 2}\right)
\times \cr & \left.
\prod_{i=3}^n \left[\sinh(\theta-\theta_j)-\sinh{i\pi B\over
2}\right]
\left[\sinh(\theta+\theta_j)-\sinh{i\pi B\over 2}\right]\right\}
}}
and we have adjusted the constant $c_n^k$ in order to absorb
the remaining constant factors.

\subsec{Operator identification.}

These equations  are somewhat similar
to those found for operators in the bulk \ref\SHG{A. Fring, G.
Mussardo,
P. Simonetti, Nucl. Phys. {\bf B}393 (1993), 413;
A. Koubek, G. Mussardo, Phys. Lett. B311 (1993), 193.}.  There are
some
essential differences though, which make the operator identification
more difficult. For instance, in the bulk, an argument based
on Lorentz invariance can be made to determine the degree
of the polynomials  $Q^k$, depending on the spin of the operators:
it is not clear how to extend this in the
boundary case.  In order to do the operator identification,
we will use below the fact that one can analytically continue the
form factors in the coupling constant $B\rightarrow 1+{2i\over\pi}
\beta_0$, leading to the so-called roaming trajectories between
minimal models \ref\roaming{Al.B. Zamolodchikov, ``Resonance
factorised scattering and roaming trajectories", ENS-LPS-335, (1991),
in Kyoto U., Yukawa Inst.}.
Then, as $\beta_0\rightarrow \infty$, at least for some simple
operators, we should
recover the form factors of the Ising model obtained
in the previous sections.
As we will see, this is not sufficient, in fact it allows to
determine only part of the polynomials $Q^k$.
Another way can be devised to
identify operators by taking the massless limit: if we know
the boundary dimension of the operator we can identify formally
the form factor series with that of the bulk.  This will allow
to determine the maximal degree of the polynomials.

A specific property of the boundary case is the
following: requiring the polynomials to be
symmetric under exchange, and invariant under reflection,
leads to the fact that they should be symmetric polynomials
in the variables $x_i=\cosh\theta_i$.  Naively, one could
then be tempted
to construct homogeneous symmetric polynomials in these
variables, but this would be wrong:
since the polynomial has to be symmetric in $e^\theta$
and $e^{-\theta}$ we can form all invariant polynomials in
these two variables.  The sum is $\cosh\theta$ but the
product, which is another symmetric polynomial in these
variables, is 1.  This means that we should not expect
the polynomials to be homogeneous.

We will not carry this study  much further here,
and contend ourselves by giving a few examples,
mostly in the case of scalar products, ie  transition form-factors
for the identity operator. Since we
choose our form factors to be normalised to 1 in the
zero particle sector (we divide by the expectation value),
we obviously have that the zero particle form factor in
simply 1.  For the two particle scalar product,
we have the equation
\eqn\polyeq{
Q^{Id}(\theta+i\pi,\theta)=1.}
where we adjusted the constant $c_2^{Id}$ such that
\eqn\cdeux{
c_2^{Id}={\cos{\pi F_b\over 2}-\cos{\pi F_a\over 2}\over 2
F_{min}(i\pi)}.
}
Obviously there are more than one way to satisfy this
equation.  Choosing $Q^{Id}(\theta_1,\theta_2)
=1$ would seem, at first sight,
to be the correct solution for the scalar products.
Following \ref\roam{C. Ahn, G. Delfino, G. Mussardo, Phys. Lett.
B317 (1993), 573.}
let us take the roaming
limit $B=1+{2i\over \pi}\beta_0$ with $\beta_0$ very
large.  Then we observe that
\eqn\roamlimit{\eqalign{
F_{min}(\theta)&\rightarrow -i
e^{-\vert \beta_0\vert/2} \sinh{\theta\over 2}
\cr
f_{ab}(\theta)^{sinh-G}&\rightarrow
e^{\vert\beta_0\vert\over 4} f_{ab}(\theta)^{Ising}.
}}
Thus the scalar product becomes that of the Ising model
provided we choose $iF_a={2\over \pi}\theta_a$.
On the other hand, if we take the massless limit and
choose boundary conditions $a$ and $b$ to be of
Neumann and Dirichlet type respectively, then from
conformal field theory the situation is described by
the insertion of an operator of dimension $h=1/8$.
Then we can compare our choice for $Q^{Id}$ with
the expansion of massless operators in the bulk having
$h=1/8$.  The conclusion, at this 2 particle level, is that
there should also be a contribution proportional to
$\sigma_1=\sum_{i=1}^2 x_i=\sum_{i=1}^2 \cosh\theta_i$.
This term would not modify the residue equation.
Thus, based on this analysis we expect
\eqn\expoly{
Q^{Id}(x_1,x_2)=A \sigma_1+1.
}
The main difficulty is that we have no means to fix
the relative normalisation $A$ and we are left with a
free parameter.

The four particle contribution is constructed using the
residue equation, we have
\eqn\shfour{
Q^{Id}(-x,x,x_3,x_4)=
P(x|x_3,x_4) Q^{Id}(x_3,x_4)
}
which allows to construct the four particle scalar product
using the property of symmetric polynomials
\foot{Here we use the notation $\sigma_k$ for the elementary
symmetric polynomials defined
by the generating function $\prod_{k} (1+t x_k)=\sum_{k} t^k
\sigma_k$.
The superscript indicates the number of indeterminates.}
\eqn\propsym{
\sigma_k^{n+2}(x,-x,x_1,...,x_n)=\sigma_k^n-x^2 \sigma_{k-2}^n.
}
The part coming from the ``1" can be fixed uniquely again
\eqn\shfours{\eqalign{
Q^{Id,1}(x_1,...,x_4)&=4 \gamma (\sigma_3\sigma_2-\gamma^2
\sigma_3)+4\gamma (\kappa_a \kappa_b-1) (-\sigma_2\sigma_1+
\gamma^2 \sigma_1)\cr &
-2i(\kappa_a-\kappa_b)(\sigma_2^2-\sigma_3 \sigma_1-4 (1+\gamma^2)
\sigma_4-\gamma^2 \sigma_1^2-2\gamma^2 \sigma_2+\gamma^4).
}}
Here $x_i=\cosh\theta_i$, $\gamma=\sinh {i\pi B\over 2}$ and
$\kappa_{a/b}=\cos{\pi F_{a/b}\over 2}$.
We observe that as expected, this part of the
solution also collapses to
that of the Ising model in the roaming limit.

For the other part, there is an ambiguity for the term
of degree $6$ coming from the kernel of the residue equation
\eqn\kernel{
\hbox{Residue }(\theta,\theta+i\pi,\theta_3,\theta_4)=0
}
and in general we will need physical arguments to choose the
correct solution.
We believe a full study of the boundary
operator content of this integrable theory could lead to a
complete solution but felt it is beyond the scope of this paper.

Finally, we dicuss the boundary form-factor for homogeneous
boundary conditions ($a=b$), in the simple case of the trace
of the stress energy tensor. The simplest guess for two particles
is
\eqn\simple{\eqalign{
Q^{\Theta}(\theta_1,\theta_2)&=\cosh\theta_1+\cosh\theta_2
\cr &= 2 \cosh({\theta_1-\theta_2\over 2})
\cosh({\theta_1+\theta_2\over 2}).
}}
Here the addition of a
constant term would spoil the residue property needed.
The normalisation is fixed by taking the roaming limit,
where the form factor should be identical with the one of the
Ising model (computed explicitely using the
mode decomposition given in appendix A).
Moreover, the massless limit reproduces the bulk form
factor up to reflection matrices.

The four particle form factor is determined using the
recurrence relation of the symmetric polynomials \eqsympol\
and we find
\eqn\polyqua{
Q^\Theta(\theta_1,\cdots,\theta_4)=
[4 \gamma (\kappa_a^2-1)(-\sigma_2 \sigma_1+\gamma^2 \sigma_1)
+4 \gamma (\sigma_2\sigma_3-\gamma^2 \sigma_3)]\sigma_1.
}
It is clear that in the roaming limit, this form factor
vanishes, as expected since in the Ising model the four
particle form factor of the energy-momentum tensor is
zero.  Moreover, assuming that
$\sigma_1$ must be factorised,
the ambiguity in this polynomial can be
fixed since by taking the massless limit we know it must have
at most degree 6.

We believe that these few examples are sufficient to
show that the method is applicable to any integrable model,
and in fact is very similar to the usual problem of finding
form factors in the bulk.
However, there is a need for a systematic way to identify
operators at the boundary and determine the free parameters
discussed above.
We now conclude by some applications.

\newsec{$P(t)$ in dissipative quantum mechanics}

The applications of our formalism to 2D statistical mechanics
problems
are somewhat obvious. We rather concentrate on quantum impurity
problems,
and discuss now  a standard problem of
dissipative quantum mechanics, the computation of ``$P(t)$'' in the
double well problem.  In the so
called ohmic regime,
the two state problem with dissipation can be mapped on a
single channel Kondo model \Chakrev, with
hamiltonian
\eqn\hamilsg{
H_\lambda=\int_{-\infty}^0 dx {1\over 2}[\Pi^2+(\partial_x\phi)^2]
+\lambda \delta (x) (S_+ e^{i\sqrt{2\pi g}\phi}+S_- e^{-i\sqrt{2\pi
g}\phi}).
}
Here, $S_\pm$ are spin $1/2$ operators, the values up and down
corresponding to the two states of the system. The dissipation
is characterized by a dimensionless number, which
can be taken to coincide with
 the conformal weight $h=g$ of the boundary operator.

The  physical quantity of interest, $P(t)$ is defined as follows:
assuming
that the spin has been fixed  in the up state up to  time  $t=0$,
$P(t)$  is its average value as a function of time after turning on
the
dissipation.
Mathematically, introducing the ground state of the theory where the
spin is not coupled to the heat bath, $|0\rangle_{\lambda=0}$, we
need
to evaluate
the one point function
\eqn\defffff{P(t)=\langle \Omega|S_z|\Omega\rangle,}
where $|\Omega\rangle=e^{-iHt}|0\rangle$,
 $|0\rangle$ is the tensor product $|0\rangle_{\lambda=0}\otimes
|+\rangle$
and $H$ is the hamiltonian \hamilsg\
with
dissipation. Equivalently, we can write
\eqn\deffffff{P(t)=\langle 0|e^{iHt}S_z
e^{-iHt}|0\rangle.}
We denote the quantity inside the bracket as $S_z(t)$ (evolution
with respect to the hamiltonian \hamilsg\ is implied).

We adress the problem in the formalism described previously.
Introducing
a complete set of eigenstates of $H_\lambda$ on the left and right
hand side
of \deffffff,
the evaluation of $P(t)$   requires the knowledge of the scalar
products
of these eigenstates with $|0\rangle_{\lambda=0}\otimes |+>$.
To diagonalize the hamiltonian, we then   use integrability of the
anisotropic Kondo model, and the  resulting massless scattering
description, involving solitons and antisolitons (of charge $\pm 1$)
and breathers.

There is a slight ambiguity in this approach, because the scattering
description
is based on an infrared picture, where the spin is always screened,
for $\lambda>0$.
For any $\lambda>0$, the spin can  be ``extracted'' from the
asymptotic states: for instance, the spin correlators follow simply
from the correlators of $\partial_x\phi$, which can be computed using
the form-factors, without any reference to the spin. At $\lambda=0$,
it is not clear ``where'' the spin exactly is in our description.
Nevertheless,
one can write axioms for the scalar products of multiparticle states
at $\lambda\neq 0$
with $|0\rangle_{\lambda=0}\otimes |+>$, and of course they are of
the
same  type as the ones written before for any pair of
couplings $\lambda,\lambda'$, so it seems our approach should work
in that case too. The effect of the spin $|+\rangle$ in $|0\rangle$
is
now simply taken into account by considering that $|0\rangle$ has a
charge equal to one, ie the only non vanishing scalar products will
be of the type
${}_\lambda^{\epsilon_{2n+1}\ldots\epsilon_1}
\langle\beta_{2n+1}\ldots\beta_1|0\rangle$, with $\sum \epsilon_i=1$.

We discuss first the case  $g=1/2$, where  the Kondo problem
essentially  decomposes into two decoupled Ising problems, with
boundary field $h\propto\lambda$ \ref\LLS{A. Leclair, F.
Lesage, H. Saleur, Phys. Rev. B54 (1996) 13597.} \foot{ In the limit
$\lambda\to\infty$,
the change of boundary conditions corresponds to the insertion of
an operator with dimension $h={1\over 8}$, twice the dimension of
the Ising spin (for arbitrary $g$, this dimension is ${g\over 4}$; in
particular it is ${1\over 4}$ at the isotropic Kondo point
\ref\A{I. Affleck, Nucl. Phys. B336 (1990) 517).}}.
At this value of $g$,
the spectrum is made of a soliton and an antisoliton, and the $R$
matrix is
\eqn\refmat{R_{+-}=R_{-+}=
i{e^\beta-iT_b\over e^\beta +iT_b},\ R_{++}=R_{--}=0.}
and
 only the two
particle form-factor of the current is non zero
\eqn\fffree{\eqalign{
&\langle
0|\partial_x\phi(x=0,t)||\beta_1\beta_2
\rangle^{\epsilon_1\epsilon_2}\propto
\delta_{\epsilon_1+\epsilon_2}\epsilon_1
e^{\beta_1/2}e^{\beta_2/2}\cr
&\left[1+R_{+-}(\beta_1)R_{+-}(\beta_2)\right]\exp\left[-it
(e^{\beta_1}+e^{\beta_2})\right].\cr}}

To proceed, we  have to compute the one point function of
$\partial_x\phi$
 by inserting  complete sets of states. Many processes will
contribute.
In the simplest, a soliton is emitted at the transition, acted on by
$\partial_x\phi$
that transforms it into another soliton, and this second soliton is
destroyed at
the second transition. The amplitude for this process is
\eqn\morei{\int{d\beta_1d\beta_2\over (2\pi)^2}
\langle 0|\beta_2\rangle_\lambda^+\ \
{}_\lambda^+\langle\beta_1|0\rangle
\ \
{}_\lambda^+\langle\beta_2\partial_x\phi(t)
|\beta_1\rangle_\lambda^+.}
This process gets ``convoluted'' with many others, where an
additional
arbitrary even number of particles is emitted at one transition,
simply ``goes through'',
and is destroyed at the other transition. This means one has to
replace
$\langle 0|\beta_2\rangle_\lambda^+\
{}_\lambda^+\langle\beta_1|0\rangle$
in \morei\ by
\eqn\normall{\langle 0|\beta_2\rangle_\lambda^+
\ {}_\lambda^+\langle\beta_1|0\rangle+\int {d\beta_3d\beta_4\over
(2\pi)^2}
\langle 0|\beta_4\beta_3\beta_2\rangle_\lambda^{+-+}\ \
{}_\lambda^{+-+}\langle\beta_4\beta_3\beta_1|0\rangle
+\ldots.}
The manipulation of this sort of terms follows from the general
principles
used in section 3.3: we  simplify them by moving contours
and using closure relations. Here, we move the $\beta_1,\beta_2$
integrals to $\hbox{ Im }\beta_1
=-i\pi$ and $\hbox{ Im }\beta_2=+i\pi$; by doing so, we do encounter
singularities
of the form-factor of $\partial_x\phi$, and kinematical poles.
Forgetting
these singularities for a while, crossing produces a
$\delta(\beta_1-\beta_2)$,
which gives a vanishing contribution because $\langle
\beta|\partial_x\phi|
\beta\rangle=0$. The kinematical poles, together with the coincident
terms in crossing, give rise to terms like $\delta(\beta_1-\beta_3)$.
By proceeding
similarly with the new integrals, we can succesively eliminate all
the integrals,
and we are left with the contribution of the poles of the form-factor
of $\partial_x\phi$. Going back to \morei\ and \normall, when we move
the contours of $\beta_1$ and $\beta_2$ integrations, there is a pole
for $\beta_1=\beta_b-{i\pi\over 2}$ and  $\beta_2=\beta_b+{i\pi\over
2}$.
For each of these poles, we are left with some integrals which we
eliminate
in the same fashion as before: all what remains at the end is
proportional to the residue of  the $\partial_x\phi$ form factor at
the
poles, ie a term $e^{-2T_bt}$, up to some numerical, time independent
constant.

There are two other types of terms contributing
to the one point fucntion of $\partial_x\phi$. In the first type, the
first transition emits two solitons and an antisoliton, a pair
soliton antisoliton
is destroyed by the $\partial_x\phi$ insertion, and the other soliton
by the second transition. In the second type, the $\partial_x\phi$
insertion
emits a pair soliton antisoliton. These processes get convoluted with
many others as before.
By moving the integrals, we find again a contribution proportional to
$e^{-2T_bt}$.

Using now the relation between $\partial_x\phi$ and the spin \LSS:
$$
S_z(t)-S_z(0)=\int_0^t \partial_x\phi(x=0,t')dt',
$$
we find immediately
\eqn\wellknown{P(t)=e^{-2T_bt},\  g={1\over 2},}
a well known result \ref\Hakim{F. Guinea, V. Hakim,
A. Muramatsu, Phys. Rev. B32(1985), 4410.}.

Remarkably, the only contribution to $P(t)$ comes from the poles
of the $R$ matrices, or equivalently of the non kinematical
poles of the transition factors. Since the contour manipulations
did not involve the explicit form of $G$ (nor the proof that two
different ways of computing
the form-factors give the same result, see section (3.3)),
it is very likely that this feature generalizes to the case
 $g\neq {1\over 2}$.

Assuming thus  that $P(t)$ is completely determined by the poles of
the transition
factors, that is the poles of the $R$ matrices, we can now write a
very
natural set of conjectures for its general form.

First, recall that for $g>{1\over 2}$, the sine-Gordon
spectrum
has no bound states. The $R$ matrix for solitons antisolitons is
independent of $g$: $R_\pm^\mp=i\tanh\left({\beta-\beta_b\over
2}-i{\pi\over 4}\right),R_\pm^\pm=0$.
We thus expect:
\eqn\qualiti{ P(t)=\sum_{n=1}^\infty a_n e^{-2nT_bt},\ g>{1\over 2}}
with $\sum a_n=1$. There are an infinite number of terms here
compared
to the $g={1\over 2}$  case
because
now all the ($U(1)$ neutral) $2n$ particle form-factors are non zero.

On the other hand, for $g<{1\over 2}$, there are bound states in the
spectrum. This means that the transition factors and the $R$ matrices
now have non trivial
poles. According to our conjecture,  these poles
contribute to  $P(t)$ by values of $e^\beta$ having a
non zero real and imaginary part, and we thus expect that $P(t)$ will
develop oscillatory components on this side of $g$, in agreement with
known numerical results, and an analytical argument in perturbation
around $g$ \Egger.
At large times, we expect the dominant contribution to come, as
usual, from
the one-breather term. Recall the one breather reflection matrix
\LSS:
\eqn\brrefm{R(\beta)=-{\tanh\left({\beta-\beta_b\over 2}-{i\pi
mg\over 4(1-g)}\right)\over
\tanh\left({\beta-\beta_b\over 2}+{i\pi mg\over 4(1-g)}\right)},}
together with  $\mu_1=2\sin{\pi g\over 2(1-g)}$,
the ratio of the breather mass parameter to the soliton mass.

This should give the leading behaviour of $P(t)$ (there is no factor
$2$
in the exponent now because a single breather can be destroyed by
$\partial_x\phi$)
\eqn\leadbeh{P(t)\propto \exp\left[-2tT_b \sin^2{\pi g\over
2(1-g)}\right]
\cos
\left[tT_b
\sin{\pi g\over (1-g)}\right].}
In particular, setting $g={1\over 2}-\epsilon$ one gets
\eqn\leadexp{P(t)\propto e^{-2T_bt}\cos\left(4T_b\pi\epsilon
t\right).}
The relation between $T_b$ and the bare parameter $\lambda$ is
$T_b\propto \lambda ^{1/1-g}$ - the coefficient of proportionality
can be found in\FLS.
Expression \leadexp\ agrees with the expansion near $g={1\over 2}$ in
\Egger. The
expression
\leadbeh\ is new as far as we know: in standard notations, we predict
the ratio of the period of oscillations to the damping factor to be
\eqn\predic{{\Omega\over \Gamma}=\cot{\pi g\over 2(1-g)}.}
It would be interesting to test this numerically (in fact, \predic\
agrees very well with measurments that appeared recently in
a paper of K. Voelker \ref\KV{K. Voelker, ``Dynamical Behavior of the
Dissipative Two-State System'', cond-mat/9712080.}).

\newsec{Conclusion}

It is clear that more work is necessary to fully determine the
transition form-factors, in particular the boundary form-factors as
well
as the scalar products, for general integrable theories. This seems
to be a purely technical matter, and we believe that all new
qualitative
aspects are well described by the examples discussed here.

To conclude, we would like to make some general remarks, illustrated
in the Ising case. From the knowledge of the transition factors,
 we have in fact obtained an expression
for any state of the theory with boundary condition $a$ in terms
of the states with boundary condition $b$. In particular, we have
$$
|0\rangle_a={}_b\langle 0|0\rangle_a\sum_{n=0}^\infty\int
\prod_{i=1}^{2n}
{d\theta_i\over 2\pi}{1\over (2n)!}
G(\theta_{2n},\ldots,\theta_1)|\theta_1,\ldots,\theta_{2n}\rangle_b
$$
Using that ground states are normalized, the scalar product of the
ground states follows:
$$
{1\over |{}_b\langle 0|0\rangle_a|^2}=\sum_{n=0}^\infty\int
\prod_{i=1}^{2n}
{d\theta_i\over 2\pi}{1\over (2n)!}
|G(\theta_{2n},\ldots,\theta_1)|^2
$$
It is not clear how useful this general expression is. For instance,
in the massless case and for free and fixed boundary conditions, this
is
nothing but the (R part) of the spin spin correlator at coincident
points in
the bulk, a manifestly divergent quantity.

It is also interesting to get back to the direct channel, and
express the boundary state in the case with two different boundary
conditions. By
matching the expressions of correlators in both channels, one finds
the
expression
$$
\eqalign{&|B\rangle_{ba}=\exp\left\{\int {d\theta\over 4\pi}
{1\over
2}\left[K_a(\theta)+K_b(\theta)\right]
Z^*(-\theta)Z^*(\theta)\right\}\cr
&\times \left[\sum_{n=0}^\infty \int \prod_{i=1}^{2n}{d\theta_i\over
2\pi}
{1\over (2n)!}G\left(\theta_1-{i\pi\over
2},\ldots,\theta_{2n}-{i\pi\over 2}\right)
Z^*(\theta_1)\ldots Z^*(\theta_{2n})\right]|0\rangle,\cr}
$$
where the poles at opposite rapidities $\theta_i=-\theta_j$ are
regulated
by taking the principal part. Going to the crossed channel, the
contours
have first to be completed by turning around the singularities from
above
or below, the residues completing the term in the exponential to give
either $K_a$
or $K_b$. The contours can then be moved by $\pm {i\pi\over 2}$  to
give integrals in terms
of $F$ or $G$, reproducing the two possible expressions for
correlators as
discussed in details for the energy in section 3.

\vskip1cm
\noindent{\bf Acknowledgments:} part of this work was carried out
at the ITP, where we benefitted from discussions with J. Cardy, A.
Leclair, P. Simonetti and A. and S. Zamolodchikov. We also thank K.
Voelker
for very useful comments, and for pointing out the agreement  of
formula \predic\ with his own results.

\listrefs

\appendix{A}{Perturbative computations.}

In this appendix we show some of the direct checks done on the
form factors discussed earlier.  If we use the action of the
introduction,
we can decompose the
fermions as
\eqn\fermdec{\eqalign{
\psi &=\int_{-\infty}^\infty d\theta [\omega e^{\theta/2} a(\theta)
e^{imx\sinh\theta
+imt\cosh\theta}+\overline{\omega} e^{\theta/2} a^\dagger (\theta)
e^{-imx\sinh\theta -imt\cosh\theta}]\cr
\overline{\psi}&=\int_{-\infty}^\infty d\theta [\overline{\omega}
 e^{-\theta/2} a(\theta)
 e^{imx\sinh\theta
+imt\cosh\theta}+\omega e^{-\theta/2} a^\dagger (\theta)
e^{-imx\sinh\theta -imt\cosh\theta}]
}}
where we have used $\omega=e^{i\pi/4}$.
We evaluate the scalar products between different states in the
simple case
where the
fields $h_a=0$ and $h_b$ are small.
The states at $t=-\infty$ are in the Hilbert space of the
theory $a$, which is at zero field, and those at
$t=+\infty$ are in the Hilbert space of theory $b$.
Then from standard perturbative arguments we obtain to second
order (which is the first non trivial one)
\eqn\secnorm{\eqalign{
{}_b\langle \theta_1\theta_2||0\rangle_a\propto& {h_b^2\over 2}
\int_0^\infty dy_1 dy_2 {\rm sgn}(y_1-y_2) \int_0^\infty d\theta_1'
d\theta_2' e^{-my_1\cosh\theta_1'-my_2 \cosh\theta_2'} \cr & \times
\left[2\cosh\left({\theta_1'\over 2}-i{\pi\over
4}\right)+2R_0^*(\theta_1')
\cosh\left({\theta_1'\over 2}+i{\pi\over 4}\right)\right] \cr &
\times \left[2\cosh\left({\theta_1'\over 2}-i{\pi\over
4}\right)+2R_0^*(\theta_1')
\cosh\left({\theta_1'\over 2}+i{\pi\over 4}\right)\right]
\cr &
\times {}_0\langle 0|a(\theta_1)a(\theta_2)a^\dagger(\theta_1')
a^\dagger(\theta_2')|0\rangle_0
}}
where $R_0$ is the reflection matrix at zero field.
The integral \secnorm\ are easily computed by using the
mode decomposition for the fermions:
\eqn\secnormi{
{}_b\langle \theta_1\theta_2||0\rangle_a\propto  h_b^2
\prod_{i=1,2} {\sinh\theta_i \over \cosh\theta_i \cosh({\theta_i\over
2}
-i{\pi\over 4})} \tanh\left({\theta_1-\theta_2\over 2}\right)
\tanh\left({\theta_1+\theta_2\over 2}\right).
}
We can do a similar computation for the four particle scalar product
to
fourth order and we find
\eqn\secnormii{
{}_b\langle \theta_1\cdots \theta_4||0\rangle_a\propto h_b^4
\prod_{i=1}^4 {\sinh\theta_i \over \cosh\theta_i \cosh({\theta_i\over
2}
-i{\pi\over 4})} \prod_{i<j}^4 \tanh\left({\theta_i-\theta_j\over
2}\right)
\tanh\left({\theta_i+\theta_j\over 2}\right).
}

\bye